\documentclass[conference]{IEEEtran}
\IEEEoverridecommandlockouts
\usepackage{cite}
\usepackage{amsmath,amssymb,amsfonts}
\usepackage{algorithmic}
\usepackage{graphicx}
\usepackage{textcomp}
\usepackage{xcolor}
\usepackage{cite}
\usepackage{balance}
\usepackage{bm}
\usepackage{subfigure}
\usepackage{algorithmic}
\usepackage{algorithm}
\usepackage{siunitx}

\newcommand\blfootnote[1]{%
	\begingroup
	\renewcommand\thefootnote{}\footnote{#1}%
	\addtocounter{footnote}{-1}%
	\endgroup}

\begin{document}
	
\bstctlcite{IEEEexample:BSTcontrol}

\title{RIS-Aided Angular-Based Hybrid Beamforming Design in mmWave Massive MIMO Systems
	\vspace{-1ex}
}

\author{\IEEEauthorblockN{Ibrahim Yildirim\textsuperscript{$\ast$,$\bullet$}, Asil Koc\textsuperscript{$\circ$}, Ertugrul Basar\textsuperscript{$\ast$}, Tho Le-Ngoc\textsuperscript{$\circ$}}
	\IEEEauthorblockA{\textsuperscript{$\ast$}CoreLab, Department of Electrical and Electronics Engineering, Koç University, Sariyer 34450, Istanbul, Turkey \\
\textsuperscript{$\bullet$}Faculty of Electrical and Electronics Engineering, Istanbul Technical University, Sariyer 34469, Istanbul, Turkey.\\
		\textsuperscript{$\circ$}Department of Electrical and Computer Engineering, McGill University, Montreal, QC, Canada \\
		Email: yildirimib@itu.edu.tr, asil.koc@mail.mcgill.ca, ebasar@ku.edu.tr,
		tho.le-ngoc@mcgill.ca
		\vspace{-3ex}}
}

\maketitle

\begin{abstract}
    	This paper proposes a reconfigurable intelligent surface (RIS)-aided and angular-based hybrid beamforming (AB-HBF) technique for the millimeter wave (mmWave) massive multiple-input multiple-output (MIMO) systems. The proposed RIS-AB-HBF architecture consists of three stages: (i) RF beamformer, (ii) baseband (BB) precoder/combiner, and (iii) RIS phase shift design. First, in order to reduce the number of RF chains and the channel estimation overhead, RF beamformers are designed based on the 3D geometry-based mmWave channel model using slow time-varying angular parameters of the channel. Second, a BB precoder/combiner is designed by exploiting the reduced-size effective channel seen from the BB stages. Then, the phase shifts of the RIS are adjusted to maximize the achievable rate of the system via the nature-inspired particle swarm optimization (PSO) algorithm. Illustrative simulation results demonstrate that the use of RISs in the AB-HBF systems has the potential to provide more promising advantages in terms of reliability and flexibility in system design.

\end{abstract}


\section{Introduction}
In envisioned sixth-generation (6G) wireless communication systems, enormously abundant spectrum at high frequencies is needed to be exploited in order to support eccentric applications which impose extreme performance requirements \cite{6G_Survey}. Although millimeter-wave (mmWave) transmission is one of the cornerstones of fifth-generation (5G) and beyond communication systems, promising technologies such as massive multiple-input multiple-output (MIMO) and reconfigurable intelligent surface (RIS) have been introduced to combat the high path loss and blockage susceptibility at these frequencies \cite{Massive_MIMO,SimRIS_Mag,Basar_Access_2019}. RISs, which are composed of a low-cost and passive reflecting elements, have recently attracted great interest by both academia and industry to improve the capacity and the transmission performance of wireless communication systems \cite{Huang_2019,Yildirim_multiRIS,Yildirim_hybrid}. 
\blfootnote{This work was supported by Scientific and Technological Research Council of Turkey (TUBITAK) under grant no 120E401.}

Beamforming (i.e., precoding in transmission and combining in reception) is an essential signal processing technique to ensure reliable communications between the transmitter and receiver.
    Although the conventional MIMO systems widely considers single-stage fully-digital beamforming, it causes two crucial challenges for the massive MIMO systems: (i) high hardware cost due to a single dedicated power-hungry radio frequency (RF) chain per antenna, (ii) large channel estimation overhead size \cite{Mass_MIMO_Precoding_Survey}.
Hybrid beamforming (HBF) is introduced as a promising solution, which interconnects analog RF-stage and digital baseband (BB)-stage via employing limited number of RF chains \cite{Mass_MIMO_Hyb_Survey}.
    Furthermore, in \cite{ASIL_ABHP_Access,ASIL_HPC_VTC,ASIL_PSO_PA_WCNC}, angular-based HBF (AB-HBF) is developed for also reducing the instantaneous channel estimation overhead size. Particularly, it builds RF-stage via slow time-varying angular information and BB-stage via reduced dimensional channel state information (CSI). 

Recently, RIS-assisted HBF studies have been introduced, in which RISs and mmWave massive MIMO systems are combined to achieve the targeted requirements for next-generation communication systems \cite{HPC_RIS_1,HPC_RIS_2,HPC_RIS_3}. In \cite{HPC_RIS_1}, the authors improved the spectral efficiency performance by jointly optimizing the RIS phase shifts and the hybrid precoder/combiner via a two-stage manifold optimization-based algorithm. While the hardware cost and energy consumption are reduced with the HBF structure proposed in \cite{HPC_RIS_1}, the direct path between the transmitter and the receiver is not taken into account. In \cite{HPC_RIS_2}, the authors considered an RIS-assisted HBF scheme for broadband MIMO system at mmWave frequency by proposing a geometric mean decomposition-based beamforming to avoid the complex bit/power allocation. 
Moreover, the authors jointly designed the RIS phase shift matrix and hybrid beamformer in \cite{HPC_RIS_3} for the RIS-assisted hybrid mmWave MIMO systems in the presence of a direct link between the transmitter and the receiver. Despite improvements in error performance and achievable rate, the existing RIS-assisted HBF studies require full CSI in hybrid beamformer design. 

\begin{figure*}[!t]
	\centering
	\includegraphics[width=1.5\columnwidth]{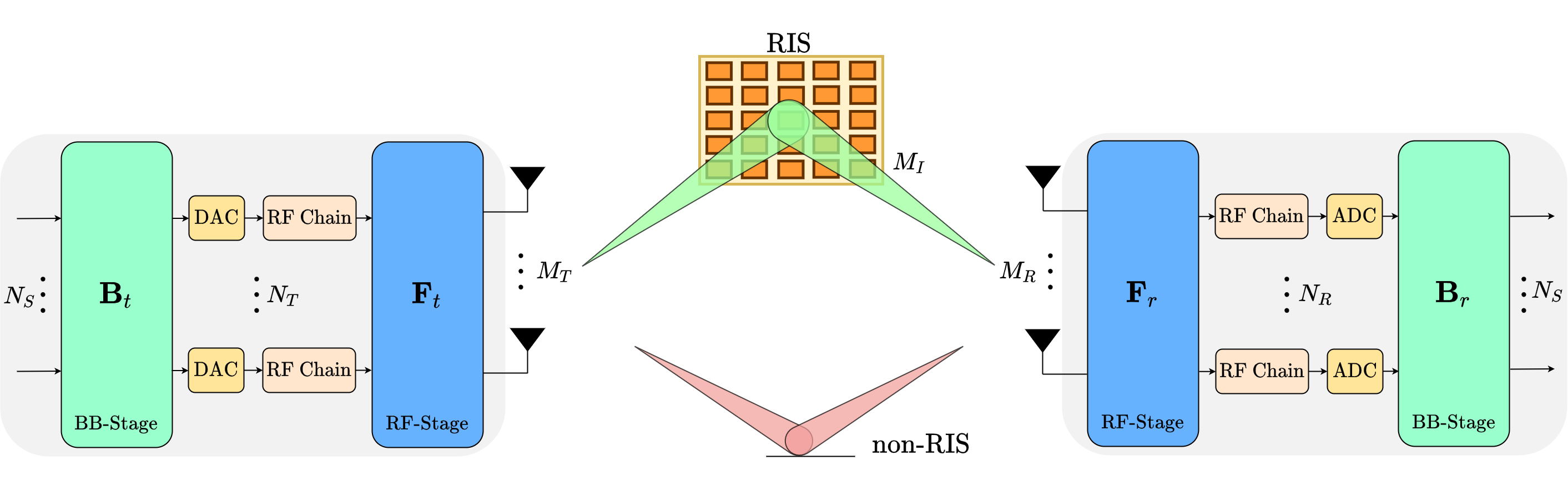}
	\vspace{-2ex}
	\caption{RIS-assisted hybrid mmWave massive MIMO system model.}
	\vspace{-2ex}
	\label{fig_1_System_Model}
\end{figure*}

In this work, we introduce an RIS-aided AB-HBF (RIS-AB-HBF) system requiring low CSI overhead for mmWave massive MIMO systems. By considering the 3D geometry-based mmWave channel model, a three-stage RIS-aided HBF system design is proposed: (i) RF beamformers, (ii) BB precoder/combiner, and (iii) RIS phase shift design.
First, the transmit/receive RF beamformers are designed by using the slow time-varying angle-of-departure/angle-of-arrival (AoD/AoA) information of the channel, while the BB precoder and combiner are constructed exploiting the reduced-size effective channel seen from the BB-stage. Then, the RIS phase reflection matrix is designed via particle swarm optimization (PSO) to maximize the achievable rate of massive MIMO systems. Our comprehensive numerical results demonstrate that the RISs significantly enhance the achievable rate performance of the AB-HBF mmWave systems by alleviating the disruptive effects of the wireless propagation environment as well as reducing the CSI overhead.\vspace{-1ex}

\section{System Model}

In this section, we present the system and channel models of the RIS-assisted HBF scheme for mmWave massive MIMO systems as well as provide their working principles.

\subsection{System Model and Problem Formulation}

We consider an RIS-assisted hybrid mmWave massive MIMO architecture shown in Fig. \ref{fig_1_System_Model}, where there is a single RIS between the transmitter/receiver pair. As shown in Fig.\ref{fig_1_System_Model}, the transmission is carried out via an RIS and non-RIS direct link between the transmitter (Tx) and the receiver (Rx). We assume that the Tx, Rx and RIS are in the form of uniform planar array (UPA). Here, the real-time controlled RIS is equipped with $M_I = M_{I_x}\times M_{I_y}$ dynamic reflecting elements that can be adjusted based on channel phases while the Tx and Rx are equipped with  $M_T = M_{T_x}\times M_{T_y}$ and $M_R = M_{R_x}\times M_{R_y}$ antennas, respectively. The number of antennas along the $x$ and $y$-axis of the corresponding UPA are respectively denoted as  $M_{i_x}$ and $M_{i_y}$ for $i \in \left\{I,T,R \right\}$. $N_S$ data streams are sent from the Tx  via digital BB precoder (${\bf B}_t$) and analog RF beamformer (${\bf F}_t$), which are connected with  $N_T$ RF chains. At the Rx, the received signal is combined via the analog RF beamformer (${\bf F}_r$) and digital BB combiner (${\bf B}_r$), which are connected with $N_R$ RF chains. Here we assume that the number of data streams is constraint as $N_S\le \min\left(N_T,N_R\right)$ and  the number of RF chains are much smaller than both the number of transmit and receive antennas to exploit the benefit of HBF systems (i.e. $N_T\ll M_T$ and  $N_R\ll M_R$). 

The precoded signal ${\bf s}\in\mathbb{C}^{M_T}$ at the Tx is expressed as:\vspace{-1ex}
\begin{equation}\label{eq_s}
\begin{aligned}
{\bf s} 
= {{\bf{F}}_t}{{\bf{B}}_t}{\bf{x}},
\end{aligned}\vspace{-1ex}
\end{equation}
where 
    ${\bf F}_t\in\mathbb{C}^{M_T\times N_{T}}$ represents the transmit RF beamformer, 
    ${{\bf B}_t\in\mathbb{C}^{N_T\times N_{S}}}$ represents the BB precoder and 
    ${\bf x}\in \mathbb{C}^{N_S}$  is the data symbol vector satisfying $\mathbb{E}\left\{{\bf x}{\bf x}^H\right\}={\bf I}_{N_s}$. 
The power constraint is also ensured for the transmitted power $P_T$ as $\mathbb{E}\big\{\big\| {\bf s} \big\|_2 ^2 \big\}\le P_T$. 

The direct channel between the Tx and Rx is represented by ${\bf H}_{TR}\in\mathbb{C}^{M_R\times M_T} $, while the RIS-assisted link consists of two components: ${\bf H}_{TI} \in\mathbb{C}^{M_I\times M_T} $ is the channel from the Tx to RIS and ${\bf H}_{IR} \in\mathbb{C}^{M_R\times M_I} $ is the channel from the RIS to Rx. 
${\bf \Theta}=\text{diag}{(e^{j\Omega_1},\dots,e^{j\Omega_{M_I}}) } \in\mathbb{C}^{M_I\times M_I}$ is  the matrix of RIS element response with $\Omega_i$ being the phase shift introduced by $i$th reflecting element of the RIS for $i=1,\dots,M_I$.
At the receiver, the signal received from the RIS-assisted and direct link is obtained as:
\begin{equation}\label{eq_y}
\begin{aligned}
{\bf{y}} =& {\bf{Hs}} + {\bf{v}} =({\bf H}_{IR} {\bf \Theta} {\bf H}_{TI}+{\bf H}_{TR}) \bf{s}+{\bf{v}}  \\
=& {{({\bf H}_{IR} {\bf \Theta} {\bf H}_{TI}+{\bf H}_{TR})}}{{\bf{F}}_t}{{\bf{B}}_t}{\bf{x}} +{\bf{v}},
\end{aligned}\vspace{-1ex}
\end{equation}
where ${\bf H}= {\bf H}_{IR} {\bf \Theta} {\bf H}_{TI}+{\bf H}_{TR}\in\mathbb{C}^{M_R\times M_T}$ stands for the end-to-end channel matrix and ${{\bf{v}}\sim\mathcal{CN}\left({\bf 0}, \sigma_v^2 {\bf I}_{M_R}\right)}$ represents the additive white Gaussian noise. 
By using \eqref{eq_s} and \eqref{eq_y}, the received signal after combining is written by:\vspace{-1ex}
\begin{equation}\label{eq_r_combined_quantized}
\begin{aligned}
\tilde{\bf y}= {\bf B}_r \bm{\mathcal{H}} {\bf{B}}_t {\bf{x}} +{\bf B}_r {\bf F}_r {\bf{v}},
\end{aligned}\vspace{-1ex}
\end{equation}
where 
	$\bm{\mathcal{H}}= {\bf F}_r {{({\bf H}_{IR} {\bf \Theta} {\bf H}_{TI}+{\bf H}_{TR})}}{{\bf{F}}_t}\in\mathbb{C}^{N_r\times N_t}$ is the effective channel matrix seen from the BB-stages,
	${\bf F}_r\in\mathbb{C}^{N_R\times M_{T}}$ represents the receive RF beamformer, 
	${\bf B}_r\in\mathbb{C}^{N_S\times N_{R}}$ denotes the BB combiner.

In this work, our main aim is to efficiently design the transmit RF beamformer and BB precoder at the Tx (i.e., ${\bf F}_t$, ${\bf B}_t$), the receive RF beamformer and BB combiner at the Rx (i.e., ${\bf F}_r$, ${\bf B}_r$), and the phase shifts at the RIS (i.e. $\bf{\Theta}$)  to maximize the achievable rate of the system as follows:\vspace{-1ex}
\begin{IEEEeqnarray}{rCl} \label{eq_capacity}
	\IEEEyesnumber
	\IEEEyessubnumber*
	\hspace{-2ex}\max_{\left\{{\bf F}_t, {\bf F}_r, {\bf B}_t, {\bf B}_r, \bf{\Theta}\right\}}&\hspace{-0.5ex}R=&\hspace{-0.5ex}{\log _2}\hspace{-0.5ex} \left|\hspace{-0.2ex}{  {{{\bf{I}}_{{N_s}}}\hspace{-1ex} +\hspace{-0.5ex} {\bf R}_{{w}}^{-1}}{\bf B}_r\bm{\mathcal{H}}{\bf B}_t{\bf B}_t^H\bm{\mathcal{H}}^H{\bf B}_r^H } \hspace{-0.3ex}\right|\hspace{-0.5ex},\\
	\hspace{-4ex}&\hspace{-30ex}\textrm{s.t.}&~\hspace{-15ex}\bm{\mathcal{H}}={\bf F}_r {\bf H F}_r\in\mathbb{C}^{N_r\times N_t},
	\label{eq_H_eff}\\
	\hspace{-4ex}& & \hspace{-14ex}\left|{\bf{F}}_t\hspace{-0.5ex}\left(m,n\right)\right|\hspace{-0.5ex}=\hspace{-0.5ex}\frac{1}{\sqrt{M_t}}, 
	\left|{\bf{F}}_r\hspace{-0.5ex}\left(m,n\right)\right|\hspace{-0.5ex}=\hspace{-0.5ex}\frac{1}{\sqrt{M_r}}, ~\forall\hspace{-0.5ex}\left(m,n\right)\label{eq_constant},\\
	\hspace{-4ex}& & \hspace{-14ex}
	\mathbb{E}\big\{ {\big\| {\bf{x}} \big\|_2^2} \big\} ={\rm{tr}}\left( {{\bf{B}}_t^H{\bf{F}}_t^H{{\bf{F}}_t}{{\bf{B}}_t}} \right) \le {P_T}\label{eq_9d},
\end{IEEEeqnarray}
where ${\bf R}_{{w}}=\sigma_v^2 {\bf B}_r {\bf F}_r {\bf F}_r^H {\bf B}_r^H$ is the covariance matrix of the noise ${\bf{w}}={\bf B}_r {\bf F}_r{\bf{v}}$. Since phase shifters are used in the implementation of RF beamformer, the constraint of the constant module is ensured as in \eqref{eq_constant}.  \vspace{-1ex}

\subsection{Channel Model}

In mmWave transmission, contrary to the conventional rich-scattering propagation environment,  only a few spatial paths are experienced due to scattering limited propagation conditions. Hence, the mmWave channel will be sparse in the angular domain \cite{Mass_MIMO_Hyb_1}.
We assume that ${\bf H}_{l}$ ($l \in  \left\{TR,TI,IR\right\}$) is a sum of the contributions of $C_l$ scattering clusters, each of which contribute $Z_{l,c}$ ($c=1,\dots,C_l$) propagation paths to the corresponding channel matrix. Each sub-channel is assumed to have $Z_l =\sum\nolimits_{c=1}^{C}Z_{l,c} $ paths in total between the terminals.

By following the UPA structure \cite{ASIL_ABHP_Access} and the 3D geometry-based mmWave channel model \cite{ChannelModels}, the channel matrix between the terminals is defined as follows:
\begin{equation}\label{eq_H_mmWave}
\begin{aligned}
{\bf H}_{l} =& \sum\limits_{c = 1}^{C_{l}} {\sum\limits_{p = 1}^{Z_{l,c}} g_{c_p}^{l} } 
{{{\bm{\phi} }} _{l,r}}\hspace{-0.25ex}\left( {{\gamma _{x,r}^{l,c_p}},{\gamma _{y,r}^{l,c_p}}} \right){\bm{\phi }}_{l,t}^H\hspace{-0.5ex}\big( {{\gamma _{x,t}^{l,c_p}},{\gamma _{y,t}^{l,c_p}}}\big)\hspace{-0.5ex} \\
=&{{\bf{\Phi }}_{l}^{r}}{\bf G}_{l}{{\bf{\Phi }}_{l}^{t}}, \: l \in  \left\{TR,TI,IR\right\},
\end{aligned}
\end{equation}
where
    $g_{c_p}^l\hspace{-0.5ex}\sim\hspace{-0.25ex}\mathcal{CN}\hspace{-0.5ex}\left(0,\frac{\beta^l}{P_l}\right)$ stands for the complex path gain of the $p^{th}$ path within the $c^{th}$ cluster of the corresponding channel with the path index of $c_p=p+\sum\nolimits_{u=1}^{c-1}Z_{l,u}$, and
    ${\bm{\phi} _{l,r}}\big( {\gamma _{x,r}^{l,c_p},\gamma _{y,r}^{l,c_p}} \big)$ 
    is the receive phase response vector while the transmit phase response vector is denoted by ${\bm{\phi} _{l,t}}\big( {\gamma _{x,t}^{l,c_p},\gamma _{y,t}^{l,c_p}} \big)$. Here, $\beta^l$ represents the path loss of the corresponding path, which is inversely proportional to the terminal distances ($d_l$) as $\beta^l \propto d_l^{-\eta}$ with $\eta$ being path loss exponent for $l \in  \left\{TR,TI,IR\right\}$.
Moreover, 
    ${\bf G}_{l}=\textrm{diag}\left(g_1^l,\cdots,g_P^l\right)\in \mathbb{C}^{Z_l\times Z_l }$ represents the diagonal matrix including  the complex path gains, 
    ${{\bf{\Phi }}_l^r}$ and ${{\bf{\Phi }}_l^t}$ denote the receive and transmit phase response matrices for the corresponding channel, respectively.

Then, the transmit and receive phase response vectors are defined as:
\begin{equation}\label{eq_phase_vector_t}
\begin{aligned}
{\bm{\phi} _{ij,t}}\hspace{-0.5ex}\left( {{\gamma_x^l, \gamma_y^l}} \right) \hspace{-0.5ex}&=\hspace{-0.75ex} \big[ {1,{e^{ j2\pi d  {{\gamma_x^l }} }}, \cdots,{e^{  j2\pi d\left( {{M_{i_x}} - 1} \right) {{\gamma_x^l}} }}} \big]^T  \vspace{-1ex}\\
&\otimes \big[ {1,{e^{ j2\pi d  {{\gamma_y^l }} }}, \cdots,{e^{  j2\pi d\left( {{M_{i_y}} - 1} \right) {{\gamma_y^l}} }}} \big]^T, 
\end{aligned}
\end{equation}
\begin{equation}\label{eq_phase_vector_r}
\begin{aligned}
{\bm{\phi} _{ij,r}}\hspace{-0.5ex}\left( {{\gamma_x^l, \gamma_y^l}} \right) \hspace{-0.5ex}&=\hspace{-0.75ex} \big[ {1,{e^{ j2\pi d  {{\gamma_x^l }} }}, \cdots,{e^{  j2\pi d\left( {{M_{j_x}} - 1} \right) {{\gamma_x^l}} }}} \big]^T\\
&\otimes \big[ {1,{e^{ j2\pi d  {{\gamma_y^l }} }}, \cdots,{e^{  j2\pi d\left( {{M_{j_y}} - 1} \right) {{\gamma_y^l}} }}} \big]^T,
\end{aligned}
\end{equation}
where $(i,j) \in  \left\{(T, R),(T, I),(I, R)\right\}$ and $d$ denotes the antenna spacing normalized by wavelength. From \eqref{eq_H_mmWave}, \eqref{eq_phase_vector_t} and \eqref{eq_phase_vector_r}, the slow time-varying transmit and receive phase response matrices for each sub-channel are respectively written by:
\begin{equation}\label{eq_Phase_Matrix}
\begin{aligned}
{{\bf{\Phi }}_l^t} \hspace{-0.5ex}=\hspace{-0.5ex} \left[\hspace{-0.5ex} {\begin{array}{*{20}{c}}
	{{\bm{\phi} _{l,t}^H}\big( {{\gamma_{x,r}^{l,1}}},{{\gamma_{y,r}^{l,1}}} \big)}\\
	{\vdots}\\
	{{\bm{\phi} _{l,t}^H}\big( {{\gamma_{x,r}^{l,P_l}}},{{\gamma_{y,r}^{l,P_l}}} \big)}
	\end{array}}
\hspace{-0.5ex}\right]\hspace{-0.5ex},~
{{\bf{\Phi }}_l^r} \hspace{-0.5ex}=\hspace{-0.5ex} \left[\hspace{-0.5ex} {\begin{array}{*{20}{c}}
	{{\bm{\phi} _{l,r}^H}\big( {{\gamma_{x,r}^{l,1}}},{{\gamma_{y,r}^{l,1}}} \big)}\\
	{\vdots}\\
	{{\bm{\phi} _{l,r}^H}\big( {{\gamma_{x,r}^{l,P_l}}},{{\gamma_{y,r}^{l,P_l}}} \big)}
	\end{array}}
\hspace{-0.5ex}\right]^H
.
\end{aligned}
\end{equation}
By considering the mean and spread parameters of AoA/AoD in Table \ref{table_Angles}, at the arrival side of each sub-channel $l$ for $l\in \left\{TR,TI,IR\right\}$, 
the coefficients reflecting the elevation AoA and azimuth AoA of the $p^{th}$ path within the $c^{th}$ cluster  are respectively defined as ${\gamma _{x,r}^{l,c_p}} = \sin \big( {{\theta _{r,c_p}^l}} \big)  \cos \big( {{\psi _{r,c_p}^l}} \big)$ 
and 
${\gamma _{y,r}^{l,c_p}} = \sin \big( {{\theta _{r,c_p}^l}} \big)\sin \big( {{\psi _{r,c_p}^l}} \big)$, where 
${\theta _{r,c_p}^l} \in \big[ {{\theta _{r,c}^l} - {\delta _{r,c}^{l,\theta}}}, {{\theta _{r,c}^l} + {\delta _{r,c}^{l,\theta}}} \big]$
and   
${\psi _{r,c_p}^l} \in \big[ {{\psi _{r,c}^l} - {\delta _{r,c}^{l,\psi}}}, {{\psi _{r,c}^l} + {\delta _{r,c}^{l,\psi}}} \big]$.
At the departure side of each sub-channel $l$ for $l\in \left\{TR,TI,IR\right\}$,
the elevation AoD and azimuth AoD of the $p^{th}$ path  within the $c^{th}$ cluster  are respectively defined as
${\gamma _{x,t}^{l,c_p}} = \sin \big( {{\theta _{t,c_p}^l}} \big)\cos \big( {{\psi _{t,c_p}^l}} \big)$
and   
${\gamma _{y,t}^{l,c_p}} = \sin \big( {{\theta _{t,c_p}^l}} \big)\sin \big( {{\psi _{t,c_p}^l}} \big)$
, where   
${\theta _{t,c_p}^l} \in \big[ {{\theta _{t,c}^l} - {\delta _{t,c}^{l,\theta}}}, {{\theta _{t,c}^l} + {\delta _{t,c}^{l,\theta}}} \big]$
and   
${\psi _{t,c_p}^l} \in \big[ {{\psi _{t,c}^l} - {\delta _{t,c}^{l,\psi}}}, {{\psi _{t,c}^l} + {\delta _{t,c}^{l,\psi}}} \big]$. 

\begin{table}[!t]
	\caption{AoA and AoD parameters of the $l^\text{th}$ channel for $l\in \left\{TR,TI,IR\right\}$.}
	\vspace{-1ex}
	\label{table_Angles}
	\centering
	\begin{tabular}{|c|c||c|c|}
		\hline
		$\theta_{r,c}^{l}$ & Mean elevation AoA & $\theta_{t,c}^{l}$ & Mean elevation AoD 	 \\ \hline
		$\delta_{r,c}^{l,\theta}$ & Elevation AoA spread & $\delta_{t,c}^{l,\theta}$ & Elevation AoD spread	 \\ \hline
		$\psi_{r,c}^{l}$ & Mean azimuth AoA & $\psi_{t,c}^{l}$ & Mean azimuth AoD		 \\ \hline
		$\delta_{r,c}^{l,\psi}$ & Azimuth AoA spread & $\delta_{t,c}^{l,\psi}$ & Azimuth AoD spread 	 \\ \hline
	\end{tabular}
	\vspace{-3ex}
\end{table}


\section{RIS-Aided Angular-Based Hybrid Beamforming}
In this section, we respectively provide the design of RF beamformers, BB precoder/combiner and phase shift matrix of the RIS for the proposed RIS-aided AB-HBF system. Our primary goal is to maximize the achievable rate of the system as well as reduce the CSI overhead size and the hardware complexity.   
From this point of view, first, the design procedures of RF beamformers is introduced based on slow-time varying angular parameters. Afterwards, BB precoder/combiner is designed by the singular value decomposition (SVD) and water filling algorithm using effective channel with reduced size. Finally, the phase shifts of the RIS, which enhances the system capacity, is obtained by applying the PSO method.  
\subsection{RF Beamformers}
In order to design RF-stage efficiently, the range of the azimuth and elevation AoA/AoD angles of the existing paths in the clusters is first required to be appropriately defined as in \cite{ASIL_HPC_VTC}. Therefore, AoA and AoD angle supports are respectively defined based on \eqref{eq_H_mmWave} as: \vspace{-1ex} 
\begin{IEEEeqnarray}{rCl}
	\IEEEyesnumber
	\IEEEyessubnumber*
	\hspace{-3.5ex}{\Upsilon_\textrm{AoA}}\hspace{-0.5ex}&=&\hspace{-0.5ex} \Big\lbrace \hspace{-0.6ex}{{{\left[ {{\gamma _{x}}\hspace{-0.25ex},\hspace{-0.25ex}{\gamma _{y}}} \right]}} \hspace{-0.5ex}=\hspace{-0.5ex} \hspace{-0.5ex}{{\left[ \sin \hspace{-0.5ex}\left( \hspace{-0.25ex}\theta \hspace{-0.25ex} \right)\hspace{-0.15ex} {\cos \hspace{-0.5ex}\left(\hspace{-0.25ex} \psi\hspace{-0.25ex}  \right)\hspace{-0.25ex},\hspace{-0.25ex} \sin \hspace{-0.5ex}\left( \hspace{-0.25ex}\theta \hspace{-0.25ex} \right)\sin \hspace{-0.5ex}\left( \hspace{-0.25ex}\psi \hspace{-0.25ex} \right)} \hspace{-0.15ex}\right]}}} \hspace{0.2ex}\Big|\hspace{0.2ex} 	\theta\hspace{-0.5ex}  \in\hspace{-0.5ex}{\bm {\theta }}_{r}\hspace{-0.25ex},\hspace{-0.25ex} \psi\hspace{-0.5ex}  \in\hspace{-0.5ex}	{\bm {\psi }}_{r}\hspace{-0.5ex}\Big\rbrace\hspace{-0.25ex},\vspace{-0.6ex}\vspace{0.5ex}\label{eq_AoA_Supp}\\
	\hspace{-3.5ex}\Upsilon_\textrm{AoD} \hspace{-0.5ex}&=&\hspace{-0.5ex} \Big\lbrace \hspace{-0.6ex}{{{\left[ {{\gamma _{x}}\hspace{-0.25ex},\hspace{-0.25ex}{\gamma _{y}}} \right]}} \hspace{-0.5ex}=\hspace{-0.5ex} {{\left[ \sin \hspace{-0.5ex}\left( \hspace{-0.25ex}\theta \hspace{-0.25ex} \right)\hspace{-0.15ex} {\cos \hspace{-0.5ex}\left(\hspace{-0.25ex} \psi\hspace{-0.25ex}  \right)\hspace{-0.25ex}, \hspace{-0.25ex}\sin \hspace{-0.5ex}\left( \hspace{-0.25ex}\theta \hspace{-0.25ex} \right)\hspace{-0.25ex}\sin \hspace{-0.5ex}\left( \hspace{-0.25ex}\psi \hspace{-0.25ex} \right)} \hspace{-0.15ex}\right]}}} \hspace{0.2ex}\Big|\hspace{0.2ex} 	\theta\hspace{-0.5ex}  \in\hspace{-0.5ex}{\bm {\theta }}_{t}\hspace{-0.25ex},\hspace{-0.25ex} \psi\hspace{-0.5ex}  \in\hspace{-0.5ex}	{\bm {\psi }}_{t}\hspace{-0.5ex}\Big\rbrace\hspace{-0.25ex},\vspace{-0.6ex}\label{eq_AoD_Supp}
\end{IEEEeqnarray}
where 
${\bm {\theta }}_{r}={\bm {\theta }}_{r}^{TI}\cup{\bm {\theta }}_{r}^{TR}$,
${\bm {\psi }}_{r}={\bm {\psi }}_{r}^{TI}\cup{\bm {\psi}}_{r}^{TR}$,
${\bm {\theta }}_{t}={\bm {\theta }}_{t}^{IR}\cup{\bm {\theta }}_{t}^{TR}$, and
${\bm {\psi }}_{t}={\bm {\psi }}_{t}^{IR}\cup{\bm {\psi }}_{t}^{TR}$. Here, the elevation and azimuth AoA supports for the corresponding sub-channels are respectively defined as: 
    ${\bm {\theta }}_{r}^l =\cup_{c=1}^{C_l}  \big[ {{\theta _{r,c}^l} - {\delta _{r,c}^{l,\theta}}}, {{\theta _{r,c}^l} + {\delta _{r,c}^{l,\theta}}} \big]$  and 
    ${\bm {\psi }}_{r}^l =\cup_{c=1}^{C_l}  \big[ {{\psi _{r,c}^l} - {\delta _{r,c}^{l,\psi}}}, {{\psi _{r,c}^l} + {\delta _{r,c}^{l,\psi}}} \big]$ for $l\in \left\{IR,TR\right\}$.
Then, the elevation and azimuth AoD supports for the corresponding sub-channels are respectively obtained as:
    ${\bm {\theta }}_{t}^l =\cup_{c=1}^{C_l}  \big[ {{\theta _{t,c}^l} - {\delta _{t,c}^{l,\theta}}}, {{\theta _{t,c}^l} + {\delta _{t,c}^{l,\theta}}} \big]$  and
    ${\bm {\psi }}_{t}^l =\cup_{c=1}^{C_l}  \big[ {{\psi _{t,c}^l} - {\delta _{t,c}^{l,\psi}}}, {{\psi _{t,c}^l} + {\delta _{t,c}^{l,\psi}}} \big]$ for $l\in \left\{TI,TR\right\}$.

By using \eqref{eq_y} and \eqref{eq_H_mmWave}, the effective channel seen from the BB-stages can be reorganized as:
\begin{equation}\label{eq_H_eff_2}
\begin{aligned}
\bm{\mathcal{H}} =&{\bf F}_r {\bf H}{\bf F}_t, \\
=&{\bf F}_r ( {{\bf{\Phi }}_{IR}^{r}}{\bf G}_{IR}{{\bf{\Phi }}_{IR}^{t}} {\bf \Theta} {{\bf{\Phi }}_{{TI}}^{r}}{\bf G}_{TI}{{\bf{\Phi }}_{TI}^{t}}  \\ 
&+{{\bf{\Phi }}_{{TR}}^{r}}{\bf G}_{TR}{{\bf{\Phi }}_{TR}^{t}})   {\bf F}_t.
\end{aligned}
\end{equation}
In order to optimize the receive (transmit) beamforming gain and take advantage of all degrees of freedom enabled by the channel, the subspace spanned by ${{\bf{\Phi }}_{IR}^r}$ and ${{\bf{\Phi }}_{TR}^r}$ (${{\bf{\Phi }}_{TI}^t}$ and ${{\bf{\Phi }}_{TR}^t}$) should be encompassed in the columns of ${\bf F}_r$ (${\bf F}_t$): 
$\textrm{Span}\left({\bf F}_r\right)
\subset \left(\textrm{Span}\left({{\bf{\Phi }}_{IR}^r}\right)\cup\textrm{Span}\left({{\bf{\Phi }}_{TR}^r}\right)\right)$ and 
$\textrm{Span}\left({\bf F}_t\right)
\subset \left(\textrm{Span}\left({{\bf{\Phi }}_{TI}^t}\right)\cup\textrm{Span}\left({{\bf{\Phi }}_{TR}^t}\right)\right)$ should be satisfied. Afterwards, the columns of the RF precoder matrix are created using the transmit steering vector 
${{\bf{a}}_t}\left( {{\gamma_x},{\gamma _y}} \right) = \frac{1}{{\sqrt {{M_t}} }}{\big[ {1,{e^{j2\pi d\gamma_x }}, \cdots ,{e^{j2\pi d\left( {{M_{T_x}} - 1} \right)\gamma_x }}} \big]^T} \otimes 
{\big[ \hspace{-0.25ex}{1,\hspace{-0.25ex}{e^{j2\pi d\gamma_y }}\hspace{-0.25ex},\hspace{-0.25ex} \cdots \hspace{-0.25ex},\hspace{-0.25ex}{e^{j2\pi d\left( {{M_{T_y}} - 1} \right)\gamma_y }}} \hspace{-0.25ex}\big]^T}\hspace{-0.5ex}\in\hspace{-0.5ex}\mathbb{C}^{M_T}$.
The quantized angle pairs are defined as follows to obtain $M_T$ orthogonal transmit steering vectors as the minimum number of angle pairs spanning the entire elevation and azimuth space \cite{ASIL_ABHP_Access,ASIL_HPC_VTC}: ${{\kappa _{x,t}^m} \hspace{-0.5ex}=\hspace{-0.5ex} \frac{2m-1}{{{M_{T_x}}}}}-1$ for $m = 1, \cdots,{M_{T_x}}$ and ${{\kappa _{y,t}^n} \hspace{-0.5ex}=\hspace{-0.5ex} \frac{2n-1}{{{M_{T_y}}}}}-1 $ for $n = 1, \cdots,{M_{T_y}}$.
Thus, while minimizing the use of the RF chain by covering the entire AoD support, the quantized angle pairs within the AoD support are written as:\vspace{-1ex}
\begin{equation}\label{eq_AOD_quantized_angles}
\left( {\kappa _{x,t}^m,\kappa _{y,t}^n} \right)~\Big|\Big.~{\gamma _x} \in {\boldsymbol{\kappa }}_{x,t}^m,{\gamma _y} \in {\boldsymbol{\kappa }}_{y,t}^n,\left( {{\gamma _x},{\gamma _y}} \right) \in \Upsilon_\textrm{AoD},\vspace{-1ex}
\end{equation}
where ${\boldsymbol{\kappa }}_{x,t}^m \hspace{-0.5ex}=\hspace{-0.5ex} \big[ {\kappa _{x,t}^m \hspace{-0.5ex}-\hspace{-0.5ex} \frac{1}{{{M_{T_x}}}},\kappa _{x,t}^m \hspace{-0.5ex}+\hspace{-0.5ex} \frac{1}{{{M_{T_x}}}}} \big]$ 
and 
${\boldsymbol{\kappa }}_{y,t}^n \hspace{-0.5ex}=\hspace{-0.5ex} \big[ {\kappa _{y,t}^n \hspace{-0.5ex}-\hspace{-0.5ex} \frac{1}{{{M_{T_y}}}},\kappa _{y,t}^n \hspace{-0.5ex}+\hspace{-0.5ex} \frac{1}{{{M_{T_y}}}}} \big]$ respectively represent the boundary of $\kappa _{x,t}^m$ and $\kappa _{y,t}^n$. Therefore, the RF precoder with $N_T$ quantized angle-pairs is constructed based on \eqref{eq_AoD_Supp} as:\vspace{-1ex}
\begin{equation}\label{eq_TX_RF}
{{\bf{F}}_t} \hspace{-0.5ex}=\hspace{-0.5ex} \big[ {{\bf{a}}_t\big( {\kappa ^{m_1}_{x,t},\kappa ^{n_1}_{y,t}} \big), \cdots, {\bf{a}}_t\big( {\kappa ^{m_{N_T}}_{x,t},\kappa ^{n_{N_T}}_{y,t}} \big)} \big]\in\mathbb{C}^{M_T\times N_T}.\vspace{-1ex}
\end{equation}

Following the similar procedure for the RF combiner ($\bf{F}_r$),  the receive steering vector is obtained as: 
${{\bf{a}}_r}\left( {{\gamma_x},{\gamma _y}} \right) = \frac{1}{{\sqrt {{M_R}} }}{\big[ {1,{e^{j2\pi d\gamma_x }}, \cdots ,{e^{j2\pi d\left( {{M_{R_x}} - 1} \right)\gamma_x }}} \big]^T} \otimes 
{\big[ \hspace{-0.25ex}{1,\hspace{-0.25ex}{e^{j2\pi d\gamma_y }}\hspace{-0.25ex},\hspace{-0.25ex} \cdots \hspace{-0.25ex},\hspace{-0.25ex}{e^{j2\pi d\left( {{M_{R_y}} - 1} \right)\gamma_y }}} \hspace{-0.25ex}\big]^T}\hspace{-0.5ex}\in\hspace{-0.5ex}\mathbb{C}^{M_R}$.
The quantized angle pairs are given by
${{\kappa _{x,r}^m} \hspace{-0.5ex}=\hspace{-0.5ex} \frac{2m-1}{{{M_{R_x}}}}}-1$ for $m = 1, \cdots,{M_{R_x}}$ and ${{\kappa _{y,r}^n} \hspace{-0.5ex}=\hspace{-0.5ex} \frac{2n-1}{{{M_{R_y}}}}}-1 $ for $n = 1, \cdots,{M_{R_y}}$.

In order to minimize the use of the RF chain in the receiver side by covering the entire AoA support, the quantized angle pairs within the AoA support are obtained as:\vspace{-1ex}
\begin{equation}\label{eq_AOD_quantized_angles}
\left( {\kappa _{x,r}^m,\kappa _{y,r}^n} \right)~\Big|\Big.~{\gamma _x} \in {\boldsymbol{\kappa }}_{x,r}^m,{\gamma _y} \in {\boldsymbol{\kappa }}_{y,r}^n,\left( {{\gamma _x},{\gamma _y}} \right) \in \Upsilon_\textrm{AoA},\vspace{-1ex}
\end{equation}
where ${\boldsymbol{\kappa }}_{x,t}^m \hspace{-0.5ex}=\hspace{-0.5ex} \big[ {\kappa _{x,t}^m \hspace{-0.5ex}-\hspace{-0.5ex} \frac{1}{{{M_{T_x}}}},\kappa _{x,t}^m \hspace{-0.5ex}+\hspace{-0.5ex} \frac{1}{{{M_{T_x}}}}} \big]$ 
and 
${\boldsymbol{\kappa }}_{y,r}^n \hspace{-0.5ex}=\hspace{-0.5ex} \big[ {\kappa _{y,r}^n \hspace{-0.5ex}-\hspace{-0.5ex} \frac{1}{{{M_{R_y}}}},\kappa _{y,r}^n \hspace{-0.5ex}+\hspace{-0.5ex} \frac{1}{{{M_{R_y}}}}} \big]$ respectively denote the boundary of $\kappa _{x,r}^m$ and $\kappa _{y,r}^n$. Finally, the RF combiner with $N_R$ quantized angle-pairs is constructed fulfilling AoA support in \eqref{eq_AoA_Supp} as:\vspace{-0.5ex}
\begin{equation}\label{eq_RX_RF}
{{\bf{F}}_r} \hspace{-0.5ex}=\hspace{-0.5ex} \big[ {{\bf{a}}_r\big( {\kappa ^{m_1}_{x,r},\kappa ^{n_1}_{y,r}} \big), \cdots, {\bf{a}}_r\big( {\kappa ^{m_{N_R}}_{x,r},\kappa ^{n_{N_R}}_{y,r}} \big)} \big]\in\mathbb{C}^{M_R\times N_R}.\vspace{-0.5ex}
\end{equation} 
It should be noted that the unitary property for RF beamformers (i.e. ${\bf F}_t^H{\bf F}_t\hspace{-0.25ex}=\hspace{-0.25ex}{\bf I}_{N_T}$, ${\bf F}_r{\bf F}_r^H\hspace{-0.25ex}=\hspace{-0.25ex}{\bf I}_{N_R}$) and the constant modulus condition provided in \eqref{eq_constant} are satisfied in the RF beamformer design stage.

\subsection{Baseband Precoder/Combiner}
After the design of the RF beamformers, this subsection focuses on the BB precoder/combiner design using the low-dimensional effective channel matrix given in \eqref{eq_H_eff_2}.  The singular value decomposition (SVD) of effective channel matrix$\bm{\mathcal{H}}$ is obtained as 
	$\bm{\mathcal{H}} \hspace{-0.5ex}=\hspace{-0.5ex} {\bf U \Sigma V}^H$,
where 
	${\bf U}\hspace{-0.5ex}\in\hspace{-0.5ex} \mathbb{C}^{N_R\times \textrm{rank}\left(\bm{\mathcal{H}}\right)}$
represents a tall unitary matrix,
	${\bf \Sigma}\hspace{-0.5ex}=\hspace{-0.5ex}\textrm{diag}\big(\sigma_1^2, \hspace{-0.5ex}\cdots\hspace{-0.5ex},\sigma_{\textrm{rank}\left(\bm{\mathcal{H}}\right)}^2\big)\in \mathbb{R}^{\textrm{rank}\left(\bm{\mathcal{H}}\right)\times \textrm{rank}\left(\bm{\mathcal{H}}\right)}$
denotes a diagonal matrix of the singular values in the descending order and
	${\bf V}\in \mathbb{C}^{N_T\times \textrm{rank}\left(\bm{\mathcal{H}}\right)}$
stands for a tall unitary matrix.
    Here, ${\bf V}$ can be partitioned as ${\bf V} = \left[{\bf V}_1, {\bf V}_2\right]$ for ${\bf V}_1\in \mathbb{C}^{N_T\times N_S}$ under $\textrm{rank}\left(\bm{\mathcal{H}}\right)\ge N_S$.
    
Similar to \cite{SU_mMIMO_Dynamic_SubArray,ASIL_HPC_VTC}, the optimal solution for the BB precoder can be represented as:\vspace{-1ex}
\begin{equation}\label{eq_TX_BB}
{\bf B}_t = {\bf V}_1{\bf \Gamma}^\frac{1}{2} \in\mathbb{C}^{N_T\times N_T},
\end{equation}
where ${\bf \Gamma}\hspace{-0.5ex}=\hspace{-0.5ex}\textrm{diag}\left(\Gamma_1,\hspace{-0.25ex}\cdots\hspace{-0.25ex},\Gamma_{N_S}\right)\hspace{-0.5ex}\in\hspace{-0.5ex}\mathbb{R}^{N_S \times N_S}$ represents a transmit power matrix with $\sum\nolimits_{n=1}^{N_S}\Gamma_n= {P_T}$ for meeting the power constraint given in \eqref{eq_9d}. By using the  water filling power control solution \cite{SU_mMIMO_OFDM}, the powers are allocated as:
\begin{equation}
{\Gamma_n} = {\left( {\mu  - \frac{\sigma_v^2}{{P_T \sigma _n^2}}} \right)^ + },{\textrm{ s.t.}}\sum\limits_{n = 1}^{N_S} {\left( {\mu  - \frac{\sigma_v^2}{{P_T \sigma _n^2}}} \right)^ + }  \hspace{-0.5ex}=\hspace{-0.25ex} P_T.\vspace{-1ex}
\end{equation}

Then, considering the given RF beamformer (${\bf F}_t$/${\bf F}_r$), phase response matrix (${\bf \Theta}$) and BB precoder (${\bf B}_t$), we strive to design the BB combiner (${\bf B}_r$) that minimize the mean-squared-error (MSE) between the transmitted and processed received signals. Based on \eqref{eq_r_combined_quantized}, the MSE expression can therefore be stated as:
\vspace{-1ex}
\begin{equation} \label{eq_MSE}
\begin{aligned}
{\bf{e}_{\rm{MSE}}} \hspace{-0.45ex} =& \hspace{-0.10ex} \mathbb{E}\big\lbrace \hspace{-0.30ex}{{{\big\| {\tilde{\bf{y}} \hspace{-0.30ex}-\hspace{-0.25ex} {\bf{x}}} \big\|}^2}}\hspace{-0.25ex} \big\rbrace
\hspace{-0.5ex}=\hspace{-0.5ex} {\rm{tr}}\big(\mathbb{E} {\big\{ {\tilde {\bf{y}}{{\tilde {\bf{y}}}^H} \hspace{-0.85ex}-\hspace{-0.5ex} {\bf{x}}{{\tilde {\bf{y}}}^H} \hspace{-0.85ex}-\hspace{-0.55ex} \tilde {\bf{y}}{{\bf{x}}^H} \hspace{-0.85ex}+\hspace{-0.45ex} {\bf{x}}{{\bf{x}}^H}} \big\}} \big)\hspace{-0.25ex}\\
=&\rm{tr}\Big(  {{\bf B}_r\bm{\mathcal{H}} {{\bf{B}}_t} {{\bf{B}}_t^H}\bm{\mathcal{H}}^H}{\bf B}_r^H
+ \hspace{-0.25ex}{\bf B}_r{\bf R}_{{q_r}}{\bf B}_r^H \hspace{-0.75ex}
 \\
 &- \hspace{-0.5ex}{{{\bf{B}}_t^H}\bm{\mathcal{H}}^H{\bf B}_r^H}\hspace{-0.25ex}+ {\bf I}_{N_s}\Big.\Big).
\end{aligned}\vspace{-1ex}
\end{equation}
For the purpose of finding the optimal solution for \eqref{eq_MSE}, the differentiation rules for the complex-valid matrices are applied. Then, the derivative of $\bf{e}_\textrm{MSE}$ with respect to BB combiner (${\bf B}_r$) is calculated as:\vspace{-1ex}
\begin{equation}
\begin{aligned}
\frac{{\partial \bf{e}_{{\rm{MS}}{{\rm{E}}{}}}  }}{{\partial {{\bf{B}}_r}}} 
&=2{\bf B}_r{\bm{\mathcal{H}}{{\bf{B}}_t}{{\bf{B}}_t^H}\bm{\mathcal{H}}^H}
-2{{{\bf{B}}_t^H}\bm{\mathcal{H}}^H}.
\end{aligned}\vspace{-1ex}
\end{equation}
Finally, the BB combiner meeting the minimum MSE criterion in \eqref{eq_MSE} can be obtained as follows:\vspace{-1ex}
\begin{equation}\label{eq_RX_BB}
{\bf{B}}_r= {\bf{B}}_t^H\bm{\mathcal{H}}^H \left( \bm{\mathcal{H}} {{\bf{B}}_t} {{\bf{B}}_t^H}\bm{\mathcal{H}}^H \right)^{-1}.\vspace{-1ex}
\end{equation}

\subsection{RIS Phase Shift Design}
In this subsection, phase shifts of the RIS elements are adjusted by optimizing the achievable rate of the system via the PSO algorithm. In PSO, each particle reconfigures its position at every step by moving through multidimensional space with its own experience and the experience of its peers towards an optimum solution by the whole swarm \cite{AI_PSO}.
Under given RF beamformers (${\bf F}_t,{\bf F}_r$) and BB precoder/combiner (${\bf B}_t$, ${\bf B}_r$), the optimization problem for maximizing the achievable rate can be formulated as: 
\begin{equation}\label{eq_OPT_2}
	\begin{aligned}
		\max_{\left\{\bf{\Theta}\right\}} ~& R\left({\bf F}_t,{\bf F}_r, {\bf B}_t, {\bf B}_r,\bf{\Theta}\right)\\
		\textrm{s.t.}~~~
		&  	{\bf \Theta}=\text{diag}{(e^{j\Omega_1},\dots,e^{j\Omega_{M_I}}) } \\
		&  	\mathbb{E}\big\{ {\big\| {\bf{x}} \big\|_2^2} \big\} ={\rm{tr}}\left( {{\bf{B}}_t^H{\bf{F}}_t^H{{\bf{F}}_t}{{\bf{B}}_t}} \right) \le {P_T}\\
		& 
		\Omega_i \in [0,2\pi ], \forall i,\vspace{-1ex}
	\end{aligned}
\end{equation}
where $R\left({\bf F}_t,{\bf F}_r, {\bf B}_t, {\bf B}_r,\bf{\Theta}\right)$ is given in \eqref{eq_capacity}. 
In our PSO solution, the phase shifts of the RIS elements ($\Omega_i$, $i\in {1,\dots,M_I}$) are modeled as the positions of the particles, and the objective function in \eqref{eq_OPT_2} is considered as the fitness function of PSO.

$N_\textrm{par}$ particles are randomly initialized with position ${\bf p} \in \mathbb{C}^{M_I}$ and velocity ${\bf v} \in \mathbb{C}^{ M_I}$. Each particle has a memory for its local best position ($ {\bf p}_{\textrm{best},i}^{(q)}$)  and the global best position (${\bf p}_{G}^{(q)} $), where the superiority of a position is evaluated by the given fitness function. Here, $ {\bf p}_{\textrm{best},i}^{(q)}$ is the best position that the $i^\textrm{th}$ particle has achieved while ${\bf p}_{G}^\textrm{(t)} $ is the best position achieved by any of the particles in the entire swarm in the $q^\textrm{th}$ iteration. For each iteration, the velocity and position of each particle are updated according to:\vspace{-1ex}
\begin{equation}\label{eq_vel_move_PSO}
    \begin{aligned}
    {\bf v}_i^{(q+1)}=&\zeta {\bf v}_i^{(q)} +u_1 \rho_1 \big( {\bf p}_{\textrm{best},i}^{(q)} - {\bf p}_{i}^{(q)}  \big) +u_2 \rho_2  \big( {\bf p}_{G}^{(q)} - {\bf p}
_{i}^{(q)}  \big),
 \\
    {\bf p}_i^{(q+1)}=& {\bf v}_i^{(q+1)}+ {\bf p}_i^{(q)},
    \end{aligned}\vspace{-1ex}
\end{equation}
where $\zeta$ is the inertia weight of velocity, $\rho_1$ and $\rho_2$ are the acceleration constants, and $u_1$ and $u_2$  are uniformly distributed random numbers in $[0,1]$.
Through the iterations, the particles move from ${\bf p}_i^{(q)}$ to ${\bf p}_i^{(q+1)}$. Finally, after $Q$ iterations, the RIS phase shift matrix is
determined at the last iteration as 
    ${\bf \Theta}=\textrm{diag}\big(\exp\big({j{\bf p}_G^{(Q)}}\big)\big)$.
The summary of the proposed RIS-aided AB-HBF system design including three stages is precisely explained in Algorithm 1.

\makeatletter
\newcommand\fs@betterruled{%
  \def\@fs@cfont{\bfseries}\let\@fs@capt\floatc@ruled
  \def\@fs@pre{\vspace*{5pt}\hrule height.8pt depth0pt \kern2pt}%
  \def\@fs@post{\kern2pt\hrule\relax}%
  \def\@fs@mid{\kern2pt\hrule\kern2pt}%
  \let\@fs@iftopcapt\iftrue}
\floatstyle{betterruled}
\restylefloat{algorithm}
\makeatother
\begin{algorithm}[t] 
    \caption{Proposed RIS-aided AB-HBF System Design}
	\begin{algorithmic}[]
		\renewcommand{\algorithmicrequire}{\textbf{Input:}}
		\renewcommand{\algorithmicensure}{\textbf{Output:}}
		\REQUIRE $\bm{\mathcal{H}}$, $\Upsilon_\textrm{AoA}$, $\Upsilon_\textrm{AoD}$, $P_T$, $Q$, $N_\textrm{par}$, $\zeta$, $\rho_{1}$, $\rho_{2}$
		\ENSURE  ${\bf F}_t$, ${\bf F}_r$, ${\bf B}_t$, ${\bf B}_r$, ${\bf\Theta}$
		\renewcommand{\algorithmicrequire}{\textbf{First stage:}}
		\STATE Build ${\bf F}_t$ and ${\bf F}_r$ via \eqref{eq_TX_RF} and \eqref{eq_RX_RF}, respectively.
		\FOR {$q = 1:Q$}
		\STATE Calculate each particle's velocity ${\bf v}_i^{(q)}$ via \eqref{eq_vel_move_PSO}.
		\STATE Calculate each particle's position ${\bf p}_i^{(q)}$ via \eqref{eq_vel_move_PSO}.
		\STATE Find each particle's achievable rate via \eqref{eq_OPT_2}.
		\STATE Update personal best $ {\bf p}_{\textrm{best},i}^{(q)}$ and global best $ {\bf p}_{{G}}^{(q)}$.
		\ENDFOR
		\STATE Set RIS phase-shift matrix as 
		    ${\bf \Theta}=\textrm{diag}\big(\exp\big({j{\bf p}_G^{(Q)}}\big)\big)$.
		\STATE Build ${\bf B}_t$ and ${\bf B}_r$ via \eqref{eq_TX_BB} and \eqref{eq_RX_BB}, respectively.
	\end{algorithmic}
\end{algorithm}

\section{Numerical Results}
\begin{figure}[!t] 
\centering
	\includegraphics[width=0.6\columnwidth]{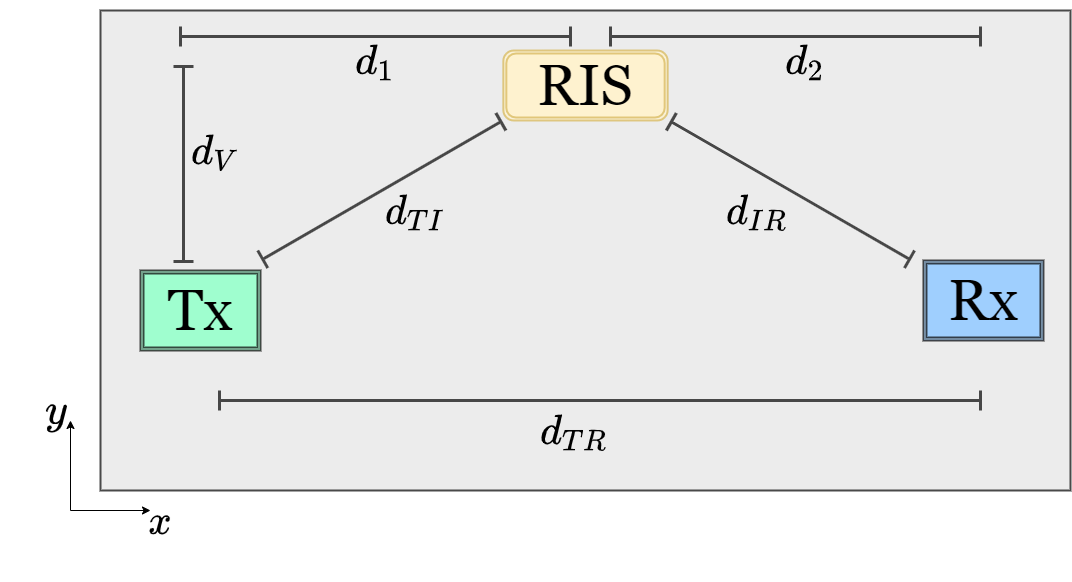}
	\vspace{-2ex}
	\caption{Top view of simulation setup for the proposed RIS-aided hybrid MIMO system.}
	\vspace{-3ex}
	\label{fig_setup} 
\end{figure}

In this section, we provide comprehensive numerical results to investigate the performance of the proposed RIS-aided AB-HBF system. Here,  
Table \ref{table_parameters} presents the common parameters for all simulations considering the geometry-based mmWave channel given in \eqref{eq_H_mmWave}. Due to the simplicity, we assume that all subchannels have $Z_{l,c}=5$ paths within a single cluster $C_l=1$ for $l\in \left\{TR,TI,IR\right\}$ and $c=1$. Moreover, the azimuth and elevation AoA/AoD spreads for all subchannels are assumed to be the same as $\delta_{i,c}^{l,j}=\ang{10}$ for $l\in \left\{TR,TI,IR\right\}$, $i=\in \left\{t,r\right\}$, $j\in \left\{\theta,\psi\right\}$ and $c=1$. The spacing between two antennas/reflectors is also $d=0.5$ due to the half-wavelength separation. The top view of the simulation setup, including the terminal distances, is also illustrated in Fig. \ref{fig_setup}. Here, the Tx-Rx channel is assumed to be nearly blocked under non-line-of-sight (NLOS) conditions, while the RIS-assisted cascade channel is assumed to have LOS conditions. 

\begin{table}[!t]
	\caption{Simulation parameters.}
	\vspace{-1ex}
	\label{table_parameters}
	\centering
	\begin{tabular}{c|c}
		\hline
		\hline
	 	\# of transmit antennas             & $M_T= 8\times 8 = 64$                         \\ \hline
	 	\# of receive antennas              & $M_R= 4\times 4 = 16$                           \\ \hline
	 	\# of RIS elements                  & $M_I= 16\times 16 = 256$                        \\ \hline
	 	\# of RF Chains                & $N_T= 6$, $N_R = 2$                        \\ \hline
        \# of data streams                  & $N_S =2$                                      \\ \hline
        \# of clusters for ${\bf H}_l$       & $C_l=1$ for $l\in \left\{TR,TI,IR\right\}$    \\ \hline
		\# of paths for ${\bf H}_l$  & $Z_{l,c}=5$ for $l\in \left\{TR,TI,IR\right\}$ \\ \hline
		Noise PSD                           & $-174$ dBm/Hz     \\ \hline
		Channel Bandwidth                   & $10$ kHz          \\ \hline
		PSO \# of iterations                & $T=200$           \\ \hline
		PSO \# of particles                 & $N_\textrm{par}=100$           \\ \hline
		Path loss exponent     & $\eta=2.3$ for LOS, $\eta=4.5$ for NLOS   \cite{mmWave_PLE}        \\ \hline
		\hspace{-3ex} Mean \hspace{-0.5ex}elevation\hspace{-0.5ex}  AoA/AoD                \hspace{-3ex}   &\hspace{-3ex}  $\theta_{i,c}^{TR} \hspace{-0.5ex}  =\hspace{-0.5ex} \ang{35}$\hspace{-0.5ex} ,\hspace{-0.5ex}  $\theta_{i,c}^{TI}\hspace{-0.5ex} =\hspace{-0.5ex} \ang{60}$, $\theta_{i,c}^{IR}=\ang{50}$, \hspace{-0.5ex} $i \hspace{-0.5ex} \in\hspace{-0.5ex}  \left\{t\hspace{-0.5ex} ,r \right\}$ \hspace{-1ex} \\ \hline 
		\hspace{-3ex}Mean \hspace{-0.5ex}azimuth\hspace{-0.5ex}  AoA/AoD                \hspace{-3ex}      & \hspace{-1ex}$\phi_{i,c}^{TR}\hspace{-0.5ex} =\hspace{-0.5ex} \ang{25}$\hspace{-0.5ex} , $\phi_{i,c}^{TI}\hspace{-0.5ex} =\hspace{-0.5ex} \ang{90}$\hspace{-0.5ex} , $\phi_{i,c}^{IR}\hspace{-0.5ex} =\hspace{-0.5ex} \ang{225}$, \hspace{-0.5ex} $i\hspace{-0.5ex}  \in\hspace{-0.5ex}  \left\{t\hspace{-0.5ex} , r \right\}$\hspace{-1ex}  \\ \hline
				\hline
	\end{tabular}
	\vspace{-3ex}
\end{table}

\begin{figure}[!t]
	\centering
	\subfigure[]
	{\includegraphics[width=0.48\columnwidth]{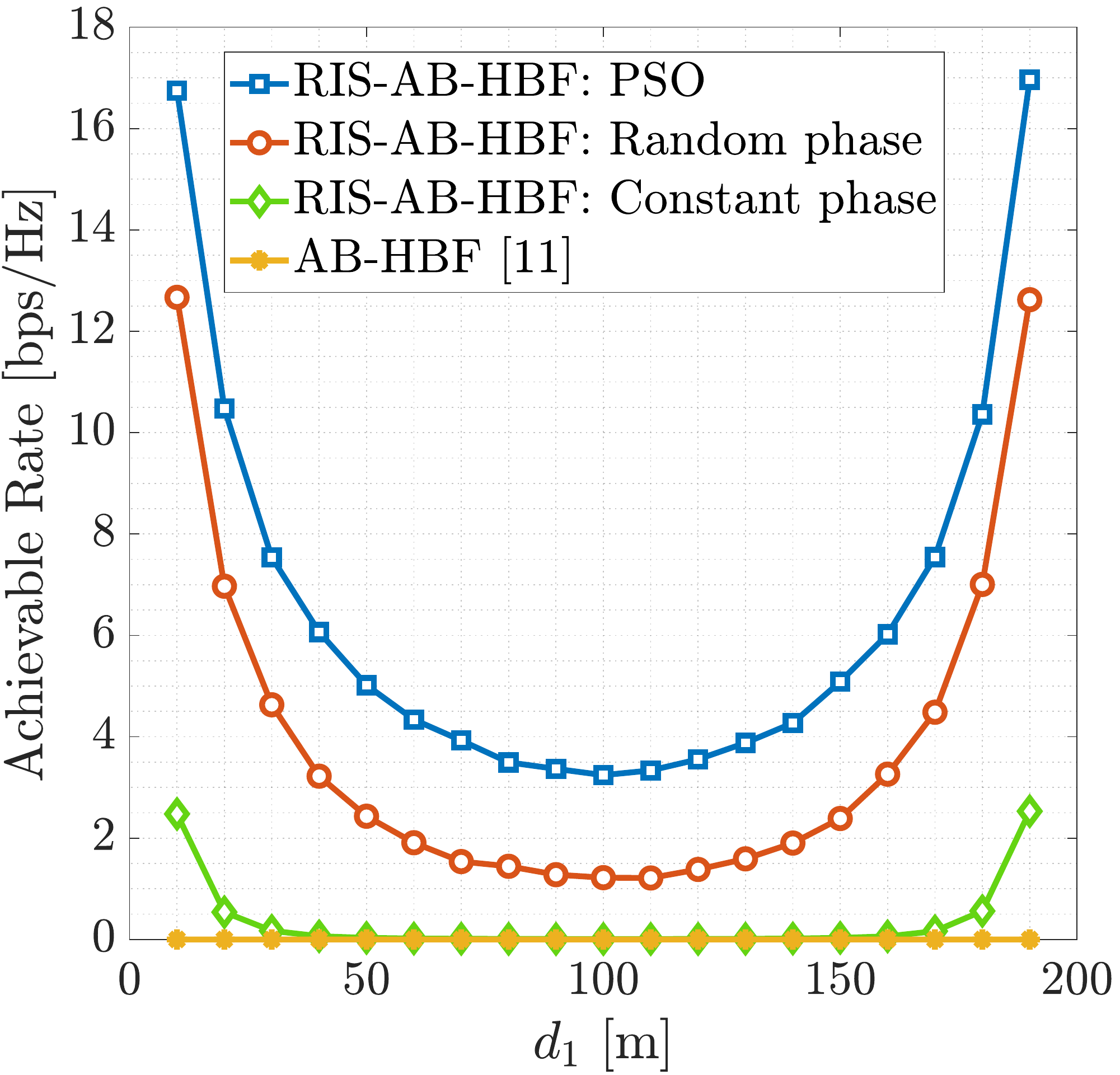}
		\label{fig_2a}}
	\subfigure[]
	{\includegraphics[width=0.47\columnwidth]{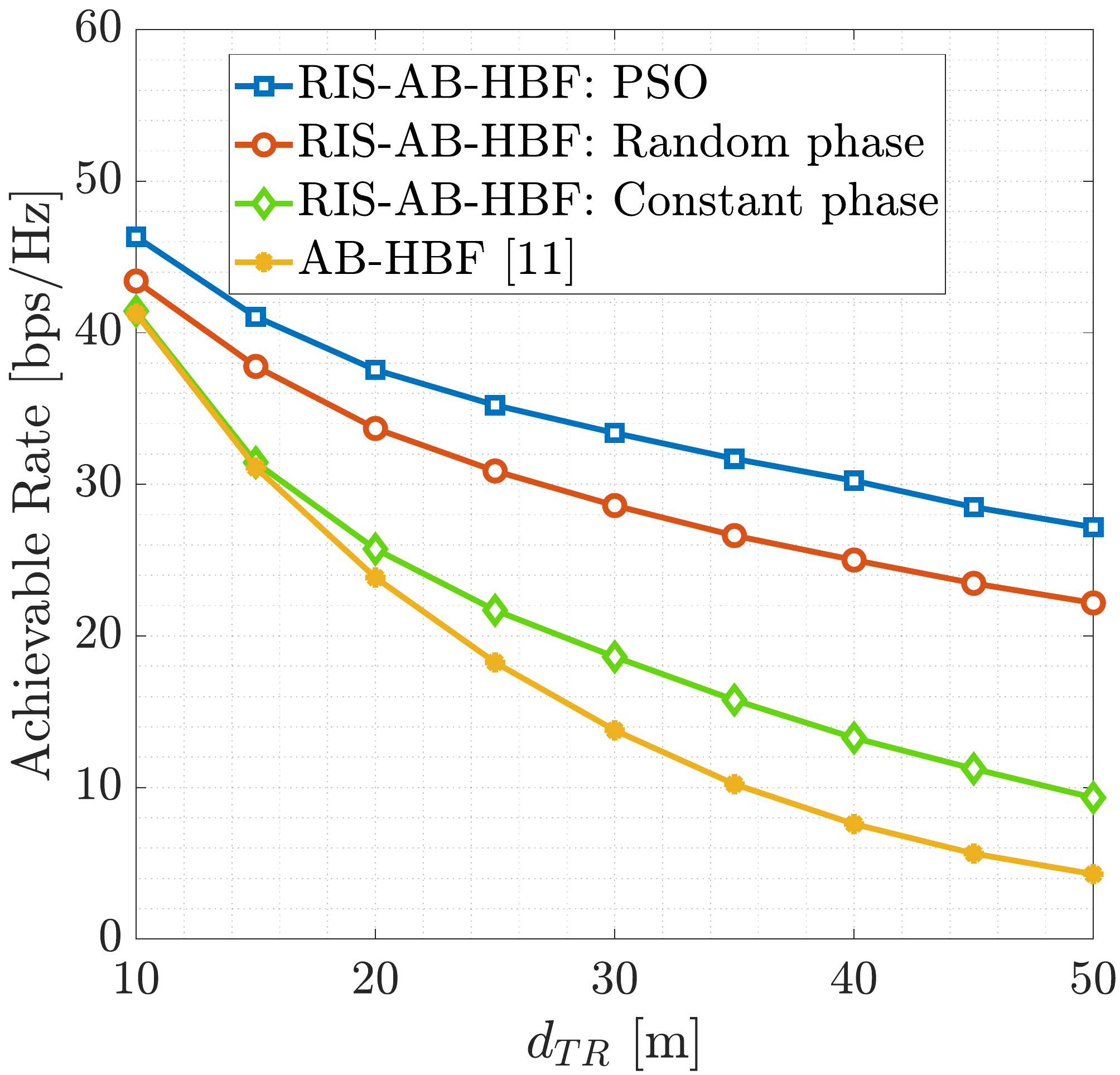}
		\label{fig_2b}}
	\vspace{-3ex}
	\caption{Comparison of achievable rate performance for AB-HBF system with or without RIS for varying (a) $d_1$ and (b) $d_{TR}$.}
	\vspace{-2ex}
	\label{fig_3} \vspace{-2ex}
\end{figure}

Fig. \ref{fig_3}(a) compares the achievable rate performance  for the AB-HBF system with or without RIS for varying Tx-RIS separation under fixed Tx and Rx positions. In this comparison, the RIS-assisted AB-HBF schemes with random and constant phase adjustment are also considered as a benchmark system for $d_{TR}=200$ and $d_V=5$ m. In this scenario, it is observed that RIS significantly improves system performance when the direct link between the Tx and Rx is nearly blocked. For $d_1=100$ m, the PSO-based RIS-assisted AB-HBF system provides $25$ dB superiority to the conventional AB-HBF system, while $4.5$ dB superiority to the random phase-adjusted RIS-AB-HBF system. Fig. \ref{fig_3}(a) also provides practical perspective for the effective positioning of the RIS. For $d_1=20$ m, the achievable rate will be measured as $10.47$ bps/Hz, while it will approximately reduce to one-third, measuring $3.24$ bps/Hz for $d_1=100$ m. In Fig. \ref{fig_3}(b), the achievable rate performance of the AB-HBF system is illustrated depending on the position of Rx changing along the $x$-axis under the fixed Tx and RIS positions. Here, the fixed distances between the Tx and RIS are given as $d_1=10$ m and $d_V=20$ m. However, doubling the $d_{TR}$ ($25$ m to $50$ m) will result in a performance loss of $22.85\%$ in the PSO-based RIS-AB-HBF system,  $28.2\%$  in the random phase RIS AB-HBF, $57.03\%$  in the constant phase RIS-AB-HBF, and $76.53\%$ in the conventional AB-HBF. From the presented results in Fig. \ref{fig_3}, we conclude that the presence of an RIS in AB-HBF systems alleviates the deteriorated channel conditions by providing an additional reliable transmission link alternative to the blocked Tx-Rx link. Thus, a remarkable increase in achievable rate performance is achieved by effectively positioning the RIS even in the absence of a LOS Tx-Rx channel.

\begin{figure}[!t]
	\centering
	\includegraphics[width=0.65\columnwidth]{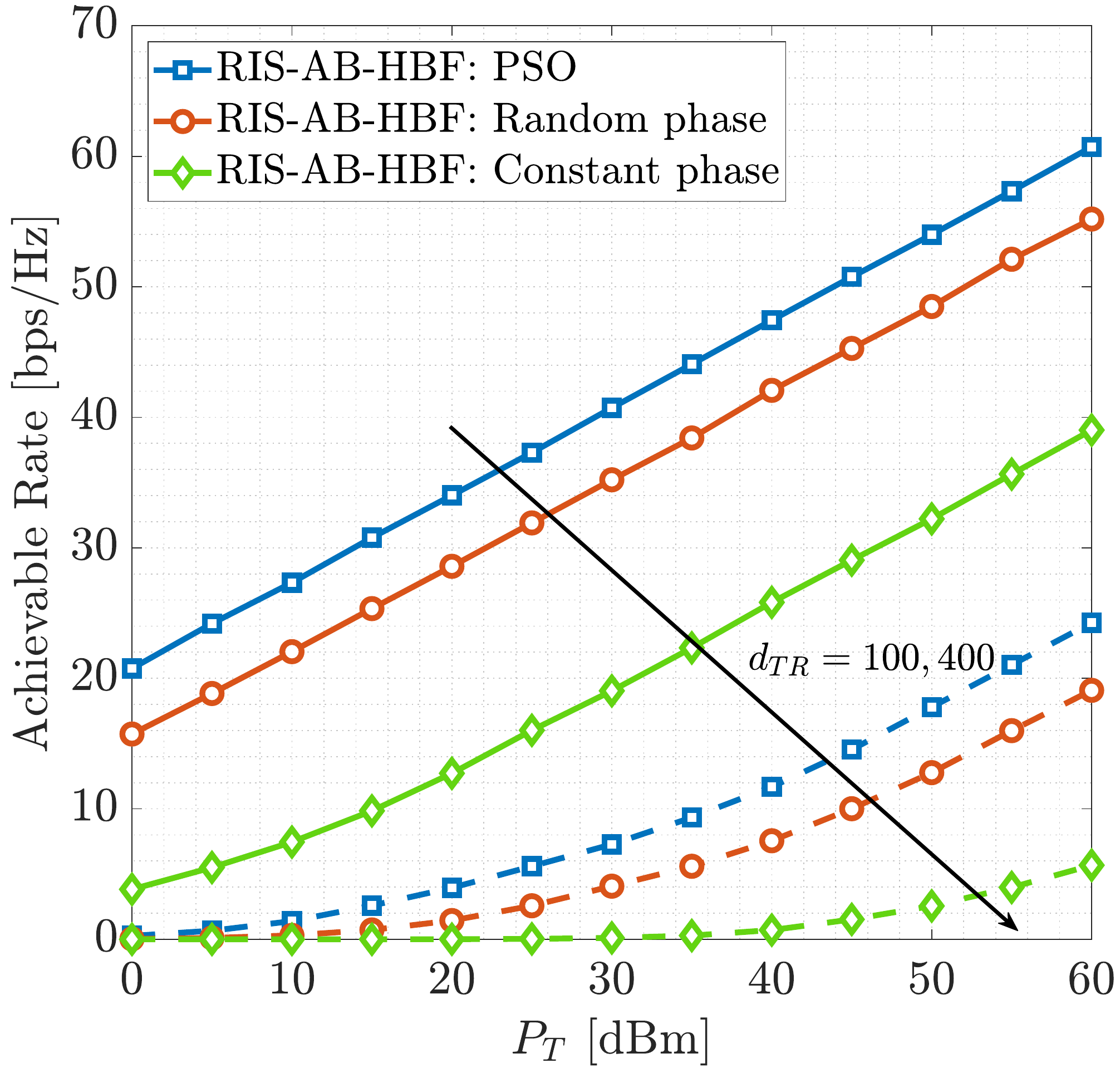}
	\vspace{-2ex}
	\caption{Achievable rate comparison of different phase adjustment methods in RIS-aided AB-HPC systems for varying $M_I$.}
	\vspace{-2ex}
	\label{fig_5} 
\end{figure}

In Fig. \ref{fig_5}, we present achievable rate comparison results for PSO-based, random and constant phase adjustment for the RIS-aided  AB-HBF system at $100$ and $400$ m $d_{TR}$ values with increasing $P_T$. Here, the distances for this setup are assumed as $d_V=5$ and $d_2=4d_1$. Fig. \ref{fig_5} gives an useful insights for the different operating modes of an RIS.  Even if the phase shifts of the RIS are set to a fixed value, a notable increase in the system performance can be achieved by providing an additional LOS path for the transmission. However, when the received signal quality is below a certain threshold value, the transmission performance can be increased by activating the PSO-assisted phase adjustment algorithm. From this point of view, the use of RISs in AB-HBF systems may has the potential to provide more promising advantages in terms of reliability and flexibility in system design.

\begin{figure}[!t]
	\centering
	\includegraphics[width=0.65\columnwidth]{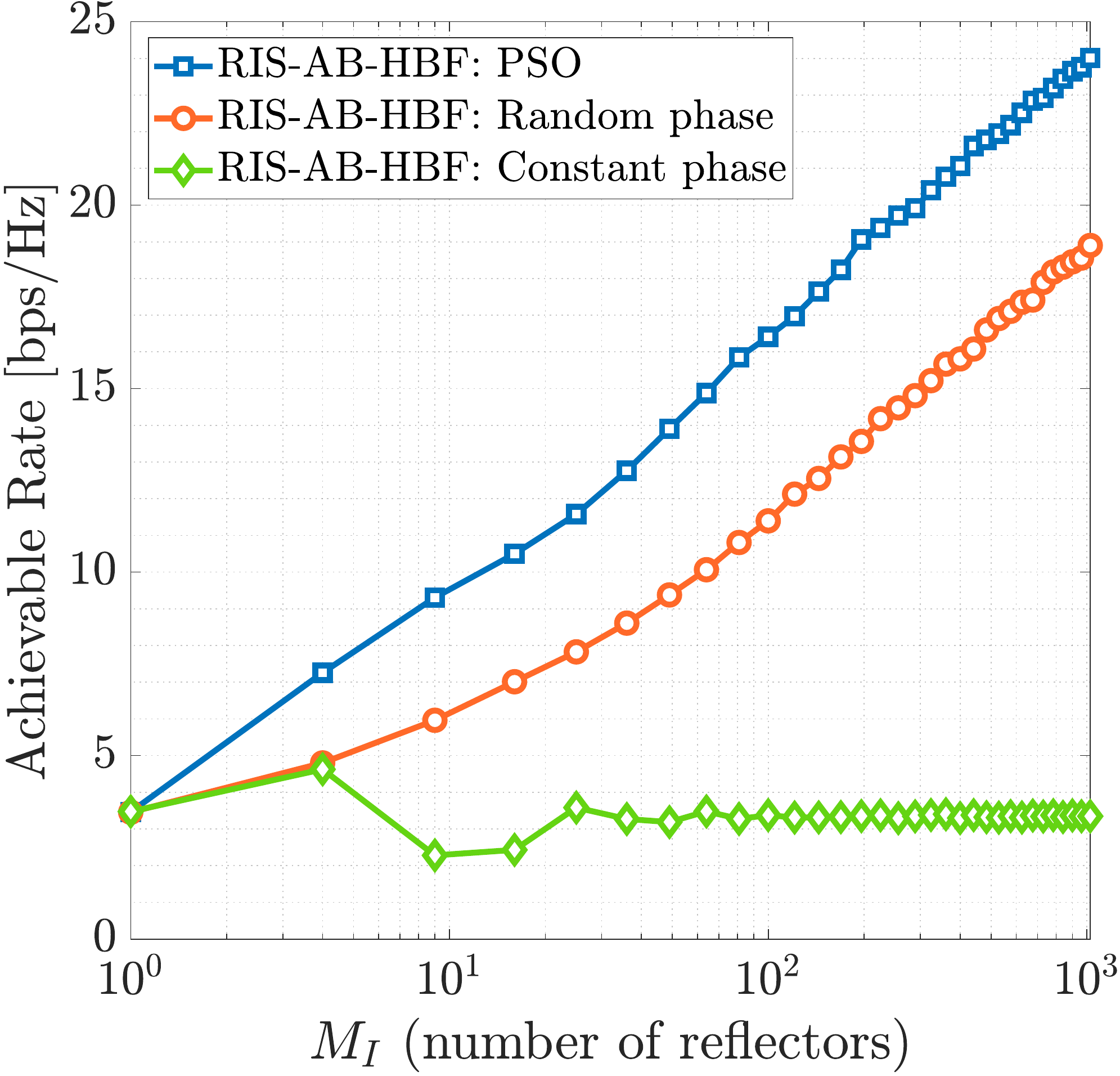}
	\vspace{-2ex}
	\caption{Achievable rate performance of the AB-HBF system with and without RIS for varying $M_I$.}
	\vspace{-2ex}
	\label{fig_4} 
\end{figure}

Finally, Fig. \ref{fig_4} demonstrates the achievable rate comparison of the AB-HPC system in the presence of an RIS under varying $M_I$ for $d_{TR}=100$, $d_1=10$, $d_2=80$ and $d_V=10$ m. As observed from Fig. \ref{fig_4}, increasing $M_I$ significantly enhance the effect of the RIS in transmission both in the case of PSO-based and random phase adjustment. While the constant phase adjustment contributes to the transmission by providing a reliable LOS path, it will become saturated at increasing $M_I$ values. Although the random phase adjustment also leads to an remarkable increase in the system performance, the effect of increasing $M_I$ is more evident for the PSO-based RIS-AB-HBF system. For instance, the PSO-based scenario provides a $3$ dB advantage over the random phase adjustment case for $M_I=10$,  while provides $4.5$ dB increase for $M_I=100$.

\section{Conclusions}
In this paper, a RIS-aided AB-HBF system has been introduced for the hybrid mmWave massive MIMO system. In this context, we have investigated a number of RIS-aided AB-HBF systems in terms of achievable rate performance to find the most efficient architecture where the RIS and HBF technologies work in harmony with each other. It has been shown via extensive numerical results that a remarkable increase in achievable rate performance is achieved by effectively positioning the RIS even in the absence of a LOS channel between the Tx and Rx. The use of RISs in the AB-HBF systems may have the potential to provide more favorable advantages in terms of reliability and flexibility in system design.  \vspace{-1ex}

\ifCLASSOPTIONcaptionsoff
\newpage
\fi
\bibliographystyle{IEEEtran}
\bibliography{bib_05_2022}
\balance

\end{document}